\documentclass[letterpaper,10pt]{article}
\usepackage{amsmath}

\RequirePackage{fix-cm}
\usepackage[left=2.5cm,right=1.5cm,top=2cm,bottom=2cm]{geometry}
\usepackage{fancyhdr,graphicx,lastpage}
\usepackage[bottom]{footmisc}
\usepackage{abstract}
\usepackage{titlesec}
\usepackage{etoolbox}
\usepackage{titling}
\usepackage{authblk}
\usepackage{indentfirst}
\usepackage{times}

\date{}

\makeatletter
  \patchcmd{\@maketitle}{\@title}{\bfseries\fontsize{12}{14.4}\selectfont\@title\vspace{-2.3mm}}{}{}
  
  \let\oldmaketitle\maketitle
  \renewcommand{\maketitle}{\oldmaketitle\vspace*{-15mm}}

  \renewenvironment{abstract}{
    \fontsize{9}{10.8}\selectfont
    \list{}{
      \setlength{\leftmargin}{0cm}%
      \setlength{\rightmargin}{\leftmargin}%
    }
    \item[\hspace{25pt}\underline{\textit{Summary}}]
    \relax}
    {\endlist}

  \fancypagestyle{plain}{
    \fancyhf{}
    
  }
  \fancyheadoffset[R,L]{-5mm}

  \renewcommand{\thesection}{}
  \renewcommand{\thesubsection}{}
  \renewcommand{\thesubsubsection}{}
  \titleformat{\section}{\bfseries\fontsize{10}{12}\selectfont\filcenter}
    {\thesection}{0em}{\uppercase}
  \titleformat{\subsection}
    {\bfseries\fontsize{10}{12}\selectfont\raggedright}
    {\thesubsection}{0em}{}
  \titleformat{\subsubsection}
    {\bfseries\fontsize{10}{12}\selectfont\raggedright}
    {\thesubsubsection}{0em}{}
  \titlespacing\subsection{0pt}{12pt plus 4pt minus 2pt}{0pt plus 2pt minus 2pt}
  \titlespacing\subsubsection{0pt}{12pt plus 4pt minus 2pt}{0pt plus 2pt minus 2pt}

  \let\oldbibliography\thebibliography
  \renewcommand{\thebibliography}[1]{\oldbibliography{#1}
  \fontsize{8}{9.6}\selectfont\setlength{\itemsep}{-3pt}} 

\makeatother

\def\be#1{\begin{equation}\label{#1}}
\def\ee{\end{equation}}
\newcommand {\ba}[2]{\be{#1}\begin{array}{#2}}
\newcommand {\ea}{\end{array} \ee}
\def\eq#1{(\ref{#1})}

\newcommand{\qq}{\,,\qquad}
\renewcommand{\=}{\stackrel{\mbox{\scriptsize def}}{=}}
\let\TS=\textstyle

\def\({\left(}
\def\){\right)}
\def\av#1{\left\langle{#1}\right\rangle}
\let\w = \omega
\def\kB{k_{\!B}}

\let\de = \delta

\def\L{{\cal L}}

\def\pt{}

\begin{document}

\title{ON HEAT TRANSFER IN A THERMALLY PERTURBED HARMONIC CHAIN}

\author[1,2]{\underline{Anton M. Krivtsov}\footnote{Email: akrivtsov@bk.ru}}
\affil[1]{Peter the Great Saint Petersburg Polytechnic University}
\affil[2]{Institute for Problems in Mechanical Engineering, Russian Academy of Sciences}
\maketitle

\begin{abstract}
Unsteady heat transfer in a harmonic chain is analyzed. Two types of thermal perturbations are considered: 1) initial instant temperature perturbation, 2) external heat supply. Closed equations describing the heat propagation are obtained and their analytical solution is constructed.
\end{abstract}

\section{INTRODUCTION}

Thermomechanical processes in ultra-pure materials differ substantially from the processes observed in a usual material. In particular, Fourier's law of heat conduction is not fulfilled in low-dimensional ultra-pure materials. This is confirmed by analytical~\cite{Bonetto 2000, Lepri 2003, Dhar 2008} and experimental~\cite{Chang-Zettl 2008, Xu 2014, Hsiao 2015} investigations.
The covariance analysis~\cite{Rieder 1967, Krivtsov 2015 DAN} allows to solve analytically the heat conduction problems for harmonic models of such materials, which are relevant to low-dimensional nanostructures. In the presented paper the closed equations describing the heat propagation in a sample system are obtained and their analytical solutions are constructed.
This paper was accepted for the XXIV ICTAM congress (21--26 August 2016, Montreal, Canada) but the author had to cancel his participation in the congress due to overlap with other commitments.


\section{THE SAMPLE MODEL}

Let us consider an infinite harmonic one-dimensional crystal (a chain of interacting particles). The dynamics equation we write in the form
\be{1}
    \dot u = v
\qq
    \dot v = \L u + b\dot W
,
\ee
where $u,\ v,\ b$ are scalar functions of time and spatial discrete variable $n$; $\L$ is a linear difference operator of the second order, $W$ is the Wiener stochastic process; the dot indicates the time derivative. The quantities $u$ and $v$ describe particle displacement and velocity, $b$ is the intensity of the random external action on the system.
The simplest variant of operator $\L$ is
\be{1a}
    \L u(n) = \w_0^2\(u(n+1) - 2u(n) + u(n-1)\vphantom{\Bigl|}\)
,
\ee
where $\w_0 \= \sqrt{C/m}$ is the frequency of the basic oscillator, $m$ is the the particle mass, $C$ is the bond stiffness. More complex operators can take into account the interaction between distant particles, elastic support and etc. The initial conditions are
\be{n3}
    u|_{t=0} = \sigma_u\rho_u \qq v|_{t=0} = \sigma_v\rho_v,
\ee
where $\rho_u$ and $\rho_v$ are random functions of $n$ with zero expectation and unit variance, $\sigma_u$ and $\sigma_v$ are deterministic functions of~$n$.

\section{DYNAMICS OF COVARIANCES}

Let us introduce covariance variables
\be{4}
    \xi(p,q) \= \av{u(p)u(q)\vphantom{\bigl|}} \qq
    \kappa(p,q) \= \av{v(p)v(q)\vphantom{\bigl|}} \qq
    \nu_1(p,q) \= \av{v(p)u(q)\vphantom{\bigl|}} \qq
    \nu_2(p,q) \= \av{u(p)v(q)\vphantom{\bigl|}},
\ee
where angular brackets stand for mathematical expectation.
Differentiation of these variables using equations of motion \eq{1} gives a closed system of differential equations for covariances:
\ba{5}{ll}
    \dot\xi  = \nu_1 + \nu_2                   \qq& \dot\nu_1 = \kappa + \L_p\xi, \\ [2mm]
    \dot\kappa = \L_q\nu_1 + \L_p\nu_2 + \beta   \qq& \dot\nu_2 = \kappa + \L_q\xi,
\ea
where $\beta(p,q) \= b(p)b(q)\de_{pq}$; $\de_{pq}=1$ for $p=q$ and it is zero otherwise.
This system can be reduced to closed systems of two equations of the second order
\be{7}
    \left\{
    \begin{array}{l}
    \ddot\xi  = 2\kappa + (\L_p + \L_q)\xi, \\ [1mm]
    \ddot\kappa = (\L_p + \L_q)\kappa + 2\L_p\L_q\xi + \dot\beta;
    \end{array}
    \right.
\qquad
    \left\{
    \begin{array}{l}
    \ddot\nu_1 = (\L_p + \L_q)\nu_1 + 2\L_p\nu_2 + \beta, \\ [1mm]
    \ddot\nu_2 = 2\L_q\nu_1 + (\L_p + \L_q)\nu_2 + \beta
    \end{array}
    \right.
\ee
or to one equation of the 4-th order
\be{8}
    \ddddot\xi  - 2(\L_p + \L_q)\ddot\xi + (\L_p - \L_q)^2\xi = 2\dot\beta.
\ee
The similar equations of the 4-th order can be obtained for variables $\kappa$, $\dot\nu_1$, $\dot\nu_2$, where the difference will be only in the right parts of the equations, which are, respectively,
\be{8b}\TS
    \(\frac\partial{\partial t} - \L_p - \L_q\)\dot\beta \qq
    \(\frac\partial{\partial t} + \L_p - \L_q\)\dot\beta \qq
    \(\frac\partial{\partial t} - \L_p + \L_q\)\dot\beta.
\ee

\section{CONTINUALIZATION}

Let us change from discrete spatial variables $p$, $q$ to variables
\be{9}\TS
    x \= a\,\frac{p+q}2 \qq n\= q-p,
\ee
where $x$ is continuum spatial variable, $n$ is discrete correlational variable, and $a$ is the lattice constant.
If the processes are relatively slow in time, and the spatial functions are relatively smooth in space, then for the case \eq{1a} the equation \eq{8} can be reduced to
\be{10}\TS
    \ddot\theta_n + \frac14c^2(\theta_{n+1}-2\theta_{n}+\theta_{n-1})'' = \dot\chi\de_n,
\ee
where
${\delta_{n} = \de_{pq}}$,
prime stands for $x$-derivative,
$c$ is the sound speed,
$\theta_n$ is the nonlocal temperature \cite{Krivtsov 2015 arXiv},
and $\chi$ is the heat supply intensity defined as
\be{11}
    c\=a\w_0 \qq
    \theta_n(x) \= (-1)^n\,\frac m{\kB}\av{v(p)\,v(q)\vphantom{\Bigl|}}\qq
    \chi \= \frac m{2\kB}\,b^2,
\ee
where $\kB$ is the Boltzmann constant. The initial conditions for equation \eq{10} are
\be{12}
    \theta_n|_{t=0} = T_0(x)\delta_n \qq
    \dot\theta_n|_{t=0} = 0 ,
\ee
where
$T_0(x)=\frac1{2\kB}m\sigma_v^2(x)$
is the initial temperature distribution,
$\sigma_u$ is accepted to be zero.
The initial conditions \eq{8} are taken after a fast transition process, which results, according to the virial theorem, in a double reduction of the initial kinetic temperature~\cite{Krivtsov 2014 DAN}.
Note that in contrast with the random initial value problem \makebox{\eq{1},~\eq{n3}} the initial value problem \makebox{\eq{10},~\eq{12}} is expressed in terms of mathematical expectations, and therefore it is a deterministic problem.

Analytical solution of problem \makebox{\eq{10},~\eq{12}} yields
\be{r5}
    T(t,x) = \frac1\pi\int_{-ct}^{ct} \frac{T_0(x-y)}{\sqrt{c^2t^2-y^2}}\,dy
    + \frac1{\pi c}\int_0^t \int_{-c\tau}^{c\tau} \ln\(\frac{c\tau+\sqrt{c^2\tau^2-y^2}}{|y|}\)\dot\chi(t-\tau,x-y)\,dy\,d\tau.
\ee
where $T(x)\equiv\theta_0(x)$ is the kinetic temperature.
Thus we have integral representation of the temperature profile in the crystal.

\section{CONCLUSIONS}

Unsteady heat conduction problems for low-dimensional nanostructures can be solved effectively using covariance analysis. The resulting solution differs substantially from solutions obtained on the basis of classical heat conduction. In particular, from the obtained solution it follows that in the considered chain any localized thermal perturbation produces the heat front, which is propagating with the sound velocity~$c$. The obtained results can be used to predict heat transfer properties in ultra-pure materials.

The author is grateful to S.N.~Gavrilov for the useful discussions.


\end{document}